\documentclass[aps,prl,reprint]{revtex4-2}

\usepackage[utf8]{inputenc}

\usepackage{comment}

\usepackage{graphicx}
\usepackage{dcolumn}
\usepackage{bm}

\usepackage{xcolor}
\usepackage{soul}
\usepackage{hyperref}
\usepackage{adjustbox}
\hypersetup{
    colorlinks=true,
    linkcolor=blue,
    filecolor=magenta,      
    urlcolor=cyan,
    pdftitle={Overleaf Example},
    pdfpagemode=FullScreen,
    }

\usepackage{natbib}
\usepackage{amsmath}

\begin{document}

\title{Upper bounds for the clock speeds of fault-tolerant distributed quantum computation using satellites to supply entangled photon pairs}
\author{Hudson Leone}
\email{leoneht0@gmail.com}
\author{S Srikara}
\email{Srikara.Shankara@student.uts.edu.au}
\author{Peter P.\ Rohde}
\email{dr.rohde@gmail.com}
\author{Simon Devitt}
\email{Simon.Devitt@uts.edu.au}
    \affiliation{%
    Center for Quantum Software and Information -- University of Technology Sydney
    }%

\date{\today}

\begin{abstract}

Despite recent advances in quantum repeater networks, entanglement distribution on a continental scale remains prohibitively difficult and resource intensive.
Using satellites to distribute maximally entangled photons (Bell pairs) between distant stations is an intriguing alternative. 
Quantum satellite networks are known to be viable for quantum key distribution, but the question of if such a network is feasible for fault tolerant distributed quantum computation (FTDQC) has so far been unaddressed.
In this paper we determine a closed form expression for the rate at which logical Bell pairs can be produced between distant surface code encoded qubits using a satellite network to supply imperfect physical Bell pairs.
With generous parameter assumptions, our results show that FTDQC with satellite networks over statewide distances (500-999 km) is possible up to a collective clock rate on the order of 1 MHz while continental (1000-4999 km) and transcontinental (5000+ km) distances run on the order of 10 kHz and 100 Hz respectively.
\end{abstract}

\maketitle

\section{Introduction}

It is well known that the Hilbert space of a quantum system grows exponentially with the number of qubits processed. 
This fact motivates research into quantum computer networking since multiple such devices working together is thought to have more computational power than the sum of its parts.
One complication of communicating quantum information though is that conventional repeater networks cannot be used to amplify a quantum signal in transit.
This is because unknown quantum information cannot be perfectly copied \cite{nocloning}.
It is therefore necessary to use ultra-reliable transmission strategies which ensure that a state is delivered with near certainty.
One well established strategy is quantum state teleportation, which uses a bipartite entangled state distributed between two distant parties as a resource for one party to communicate a qubit of information to the other \cite{teleportation}.

The quantum repeater was the first technology proposed for long distance entanglement distribution \cite{repeaterReview} but is presently considered infeasible for continental distances since it requires expensive repeater stations every 100 km or so \cite{repeater1, repeater2, repeater3, repeater4,repeater5,repeater6,repeater7}.
A more viable alternative may be to use satellites to distribute maximally entangled photons (Bell pairs) between distant ground stations.
The seminal Quantum Experiments at Space Scale (QUESS) project demonstrated that a quantum satellite today can distribute Bell pairs over distances of 1200 km at a rate around one kilohertz \cite{Mozi}.
Additional theoretic work indicates that satellite networks perform suitably well for quantum key distribution (QKD) \cite{Bonato_2009, Khatri_2021}.
Unlike QKD however, a fault-tolerant distributed quantum computation requires a continuous, high-volume supply of high-fidelity Bell pairs.

\begin{figure}[ht]
    \includegraphics[scale=0.5]{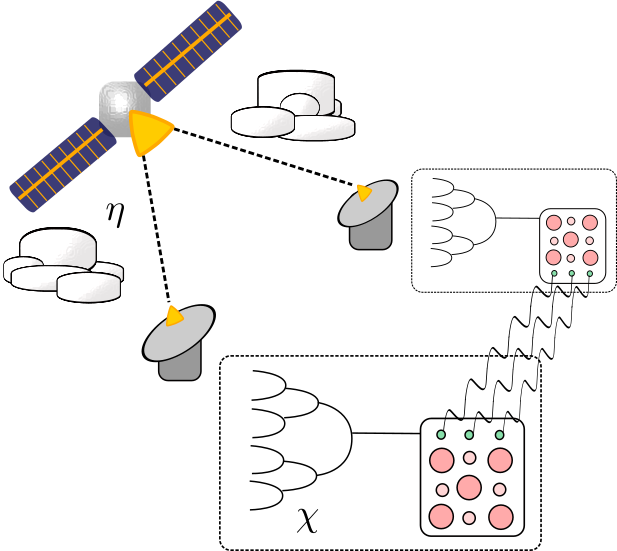}
    \caption{\label{fig:setup} A basic schematic of our scenario: A satellite in a constellation distributes entangled photon pairs between two distant ground stations. We assume an average double down-link attenuation $\eta$ due to atmospheric and free-space effects with ideal weather conditions, and we assume perfect photon capture, and conversion at the ground stations.
    To compensate for entanglement degradation, a non-deterministic recursion protocol is used to purify $\chi$ pairs into a single pair of sufficient quality.
    Finally, the purified pair is used to implement a fault-tolerant lattice surgery operation to create a logical Bell pair between the code patches.
    }
\end{figure}

In this paper, we determine a closed form expression for the rate at which satellites can produce logical Bell pairs between distant error-corrected qubits.
This in turn is the rate at which fault-tolerant quantum state teleportation can be performed, and equivalently is the clock speed of the distributed quantum computer.
Using generous parameter assumptions, we find that this clock speed is upper bounded on the orders of 1 MHz for state distances, 10 kHz for continental distances, and 100 Hz for transcontinental distances.
Since the power available to a satellite naturally limits the rate at which entanglement can be supplied, this suggests long-term scalability issues for satellite based FTDQC.
The choice of computational problem is incidental to our consideration of resource estimation, but for the sake of completeness, we choose to consider RSA public key factorization using Shor's algorithm.
We chose the surface code as our logical qubit encoding due to its high physical error tolerance, inexpensive two qubit operations via lattice surgery \cite{Horsman_2012}, and because it is the preferred encoding method for current hardware \cite{fowler2012surface} \cite{Krinner22}.
Similarly, we believe this choice of encoding is incidental to our final result since most of the Bell pairs are expended -- not on the codes themselves -- but on free-space attenuation and atmospheric effects.
Based on the logical error tolerance required for this algorithm, we compute the required rate of high-fidelity Bell pairs needed to perform a lattice surgery operation (the fault tolerant parity check that's used to produce the logical Bell state).
Working backwards, we calculate the rate of lower fidelity Bell pairs incident on the ground needed to perform entanglement purification with sufficiently high confidence.
Finally, we incorporate free space and atmospheric attenuation to estimate the rate of Bell pairs that a given satellite must be able to produce. Fig. \ref{fig:setup} shows a diagrammatic overview of our scenario.
We convert this rate to the required satellite power assuming the use of the brightest available Bell source, and with an estimate of the maximum power of a commercially available satellite.
Finally, we calculate the creation rate of the logical Bell state.

\section{Introducing the surface code and Lattice surgery}

\begin{figure*}[ht] 
    \includegraphics[scale=0.35]{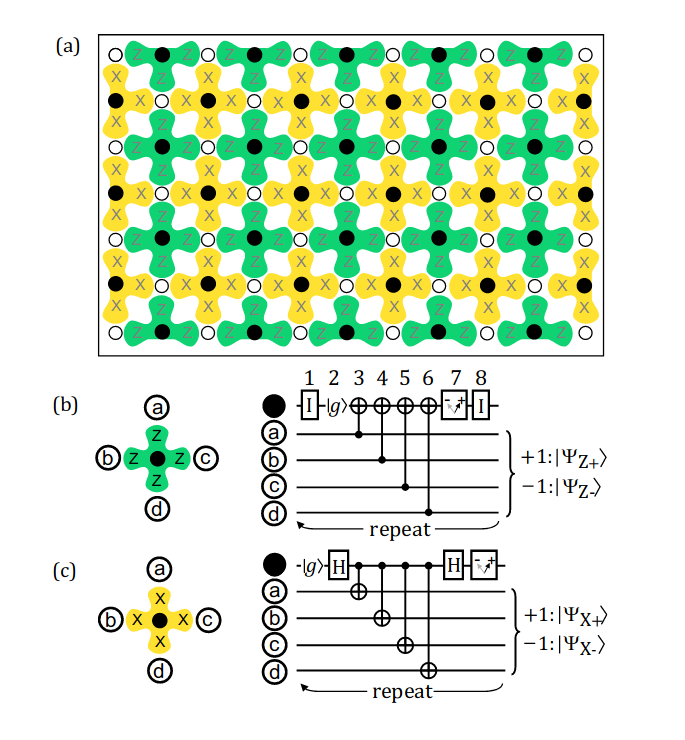}
    \caption{\label{fig:surface_code}
    (a): A top-down schematic of the surface code.
    The white (black) circles are the data (syndrome) qubits and the green (yellow) clover structures are the vertex (face) plaquettes.
    (b, c): 
    Syndrome extraction circuits for the two types of stabilizer generators for the surface code respectively.
    Figure reproduced from Horsman et.\ al.\ \cite{fowler2012surface}}
\end{figure*}

\begin{figure}[ht]
    \includegraphics[scale=0.55]{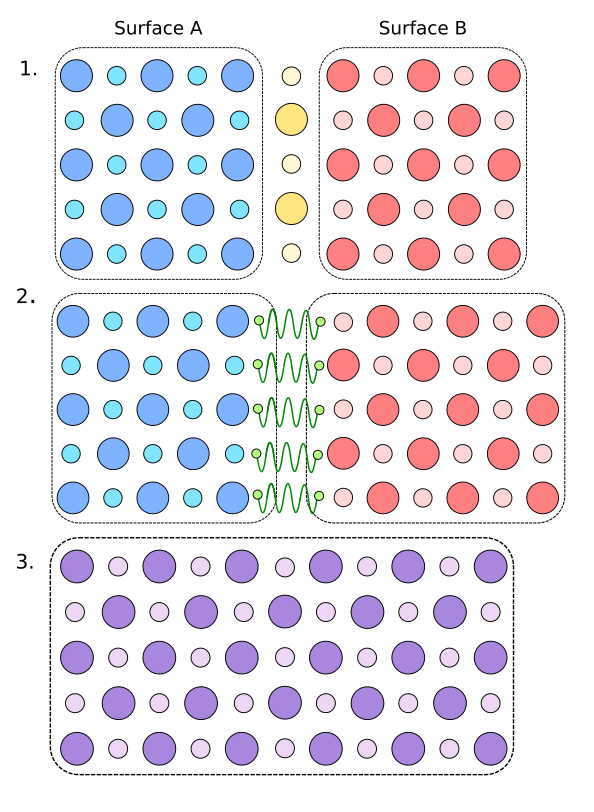}
    \caption{\label{fig:merge} 
    An illustration of lattice surgery between two spatially separated surface codes $A$ and $B$. 1) A buffer of syndrome and data qubits is initialized between the two surfaces for continuity of the qubit pattern. 2) The buffer qubits are merged into surface $B$ and Bell pairs are delivered for each qubit pair on the boundary.
    3) The syndrome extraction cycle of the code proceeds as normal. 
    This joins the two patches into a single code.
    }
\end{figure}

\begin{figure}[ht]
    \includegraphics[scale=0.65]{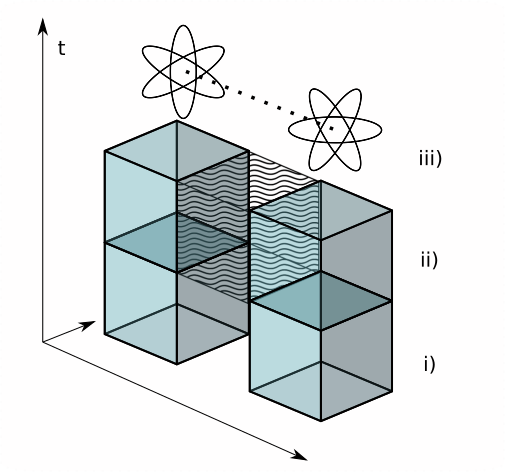}
    \caption{\label{fig:spacetime} 
    A simplified space-time diagram for generating a logical Bell state between two surface codes. Each solid cube represents $D$ cycles of syndrome extraction where $D$ is the code distance.
    i) Initialization of surface code qubits in the $|00\rangle_L$ state. 
    ii) Lattice merging with shared entanglement represented by the wavy cube connecting the two code patches.
    iii) Lattice splitting, which requires a single round of syndrome extraction.
    }
\end{figure}

We begin with a brief overview of the surface code and the lattice surgery operation.
For a more complete description, we advise the reader to consult \cite{Horsman_2012}.
A surface code is a stabilizer code consisting of two types of qubits (differing only in their respective function) arranged in a checkerboard pattern (Fig.\ \ref{fig:surface_code}a).
Each lattice can encode one logical qubit, and the larger the lattice is, the less likely the logical qubit is to suffer an undetectable error.
The \textit{data qubits} of the lattice encode the quantum information of the logical qubit while the \textit{syndrome qubits} are used to periodically measure the stabilizer generators of the code.
This process of measuring the stabilizer generators is also known as \textit{syndrome extraction}.
By performing syndrome extraction, one gains partial information about any accumulated errors on the data qubits, which allows one to correct the encoded state with high probability.
There are two types of syndrome extraction for the surface code, each of which can be represented as a five qubit circuit (Fig.\ \ref{fig:surface_code}b, c).
These circuits differ only with respect to their measurement bases, and are shown diagrammatically as clover-like tiles that cover the lattice.
Importantly, all syndrome extraction circuits can be run in parallel meaning the time it takes to implement a single \textit{syndrome extraction cycle} is $6T$ -- the depth of the circuit multiplied by the average gate time of the architecture.


We couple two surface code qubits using a technique called \textit{lattice surgery}.
This is done by performing a syndrome extraction cycle over two code patches as if they were merged together (Fig.\ \ref{fig:merge}).
We uncouple the state by measuring the qubits along the seam connecting the patches.
This \textit{merging} and \textit{splitting} is equivalent to performing an $XX$ parity measurement between the two logical qubits.
When both codes are initialised as $|0\rangle_L$, the resulting state is a maximally entangled Bell pair.
Notably, lattice surgery can be performed on surface codes that are \textit{not directly adjacent}.
Distributed Bell pairs can be used to teleport the two-qubit gates needed for the syndrome extraction circuits along the seam of the code (Fig.\ \ref{fig:merge}).
What this means is that two parties can establish error-corrected entanglement between a pair of surface codes provided they have sufficiently many physical Bell pairs to perform the lattice surgery operation.

How many entangled pairs are required for lattice surgery, and what is the rate at which they're required?
These will depend on the size of the surface codes.
It was previously mentioned that larger codes are more fault-tolerant than smaller ones. This is quantified with \textit{code distance} which for a square lattice is roughly equal to the number of qubits on either edge of the lattice.
For two codes of distance $D$, each lattice surgery would ideally require $D$ Bell pairs, as indicated in Fig.\ \ref{fig:merge}.
One subtlety that must be addressed however is that measurement errors make syndrome extraction unreliable.
In practice, we require $D$ syndrome extraction cycles for each logical operation of the surface codes.
This means that $D^2$ Bell pairs are required in total.
Given that the time to implement one round of syndrome extraction is $6T$, we require $6TD$ units of time for the $D$ rounds of fault-tolerant surgery.
The corresponding production rate of logical pairs is therefore

\begin{equation}
    R_{LP} \equiv \frac{1}{6TD}
\end{equation}

Which, when multiplied by the required number of Bell pairs, gives us the rate of ideal physical pairs needed to sustain logical pair production at a rate of $R_{LP}$:

\begin{equation} \label{eq:R_IP}
    R_{IP} = D^2 R_{LP} = \frac{D}{6T}
\end{equation}

We advise the reader at this stage that table \ref{tab:constants} catalogues the definitions of constants used throughout the paper for easy reference.




\section{Establishing code distance}

In general, we want to reduce the code distance in order to minimize the number of physical pairs needed for lattice surgery.
We therefore aim to find the smallest possible code distance that is still sufficiently large to protect logical qubits in a practical instance of quantum computing.
This first requires an understanding of how the code distance relates to the error rate of the encoded qubit.
That relationship is given:


\begin{equation} \label{eq:one_surface}
    P_L = \alpha(\beta p )^{\frac{D+1}{2}}
\end{equation}

Where $p$ is the physical error rate, and $P_L$ is the logical error rate \textit{per syndrome extraction} cycle.
Devitt et.\ al.\ \cite{sneakernet} propose parameter values $\alpha = 0.3$ and $\beta = 70$ based on their numerical data which we will adopt as well.
Recall that one round of syndrome extraction is unreliable due to measurement errors, and that all surface code operations require $D$ rounds of syndrome extraction for fault-tolerant execution.
Assuming that $P_L$ is small, we approximate the overall success rate for a surface code operation to the first order as:

\begin{equation}
(1-P_L)^D \approx (1-DP_L)
\end{equation}

The process of generating a logical pair from start to finish requires \textit{4 surface code operations} in total.
This is most easily visualised with a space-time diagram as shown in fig \ref{fig:spacetime}.
Here, each solid cube represents the $D$ counts of syndrome extraction required for each operation.
The two disjoint cubes at the base represent the preparation of the $|00\rangle_L$ state.
The middle two cubes with the wavy cube in-between represent a lattice surgery with shared entanglement on the code boundary.
Finally, a splitting operation is done which
(unlike the other operations) requires only one syndrome-cycle and is therefore considered negligible.
The total success rate of the logical pair preparation is therefore




\begin{equation}
    (1-DP_L)^4 \approx 1-4DP_L
\end{equation}

Which means the failure rate of logical pair production as a function of code distance is

\begin{equation}
    P_{LB} \equiv 4DP_L = 4D\alpha(\beta p )^{\frac{D+1}{2}}
\end{equation}

To determine a reasonable value for $D$, we need to substitute $P_{LB}$ with the tolerable error rate for some non-trivial quantum circuit.
For our calculations, we consider Shor's prime-factorization algorithm for RSA public key breaking.
This is a well established benchmark with implications in cyber-security, though itself has little bearing in the context of this paper.
The circuit we consider for implementing this factorization is given by Beauregard \cite{Beauregard}, which in the case of 2048 bit factorization can tolerate a logical error rate of $4.28\times 10^{-21}$ \cite{quantumresources}.
Given $P_{LB} = 4.28\times 10^{-21}$, and assuming a physical error rate $p=0.001$, we find that $D \approx 37$.

\section{Accounting for purification}

An entangled pair distributed through a noisy channel naturally loses some of its entanglement through decoherence.
The same is true for an entangled photon pair passing through the atmosphere.
This decay must be rectified with an entanglement distillation (purification) protocol which takes a number of low-quality pairs and produces a high-quality pair with some overall success probability using local operations and classical communications \cite{Dur_2007}.
For lattice surgery to succeed with satellite based entanglement communication, we require an additional resource overhead to account for the losses due to purification -- not only because $n$ pairs are converted into one, but also because the protocol is non-deterministic.
On this latter point, we require the entire purification process to succeed with a rate $S$ close to one.
This means that given some initial quantity of imperfect pairs, we need to be sure up to confidence $S$ that we can purify the required number of pairs needed for lattice surgery to succeed.
This is done by \textit{circuit multiplexing} wherein many instances of a protocol are performed in parallel in order to improve the probability of a successful outcome.
Circuit multiplexing is a common technique in linear-optical quantum computing where most operations are non-deterministic.



Note that entanglement purification cannot produce maximally entangled pairs in practice because of experimental uncertainty.
For this reason, we define an \textit{ideal pair} to be a state $\rho_{AB}$ such that its \textit{fidelity} with respect to a maximally entangled pair (for example $|\phi^+\rangle \langle \phi^+|$ is close to one.

\begin{equation}
    F(\rho, |\phi^+\rangle \langle \phi^+|) \geq 1 - \epsilon
\end{equation}

\begin{equation}
    F(\rho, \sigma) = \left(\mathrm{Tr}(\sqrt{\sqrt{\sigma} \rho \sqrt{\sigma}})\right)^2
\end{equation}

Let the \textit{purification factor} $\chi$ be the number of non-ideal pairs required to generate one ideal pair with confidence $S$.
The rate of physical pairs required for lattice surgery at rate $R_{LP}$ given some purification process is therefore

\begin{equation} \label{eq:R_IP+P}
    R_{IP+P} = R_{IP} \; \chi
\end{equation}



In general, the purification factor will depend on the initial state $\rho_{in}$, the required output fidelity $F_{id}$, the required success rate $S$, and the choice of purification protocol.
Determining optimal purification protocols for arbitrary mixed states remains an open challenge so
for the sake of argument, we choose to consider the well-established \textit{parity-check recurrence protocol} by Bennett et.\ al.\ \cite{Bennett_1996}.
In the most optimistic scenario, this protocol takes two pairs of the form:

\begin{equation}
    \rho_{0} = F|\phi^+\rangle \langle \phi^+| + (1-F)|\phi^- \rangle \langle \phi^-|
\end{equation}

and with probability $F^2 + (1-F)^2$ returns a state of the same form with a new fidelity of



\begin{equation}
\label{eqn:purfid}
    f(F) = \frac{F^2}{F^2 + (1-F)^2}
\end{equation}

The reason the protocol is said to be a \textit{recurrence protocol} is because the output pairs can be used as inputs for a subsequent round of purification.
In this way, the protocol can be repeated until the target fidelity is reached.
Let $N$ be the minimum number of purification rounds needed to reach the threshold fidelity $F_{\textrm{id}}$ from an initial fidelity of $F_0$.

Given the required number of purification rounds, 
what is the overall likelihood of the multi-round protocol succeeding?
Let $p(F)$ denote the success probability of a single $2 \rightarrow 1$ purification block

\begin{equation}
    p(F) = F^2 + (1-F)^2
\end{equation}

and let $F_0, F_1, \dots, F_{N-1}$ denote the input pair fidelities for the respective purification rounds.
Note that the $k$th purification round contains $2^{k-1}$ many $2 \rightarrow 1$ purification blocks that each need to succeed in order for the next round to go ahead.
The overall success probability for the protocol is therefore

\begin{equation}
    P \equiv p(F_0)^{2^{N-1}} \times p(F_1)^{2^{N-2}} \times \cdots \times p(F_{N-1})^{2^0})
\end{equation}

How many times must we multiplex a protocol with overall success probability $P$ in order to guarantee one success with a confidence of at least $S$?
Let $\mathcal{B}(P, k)$ be a binomial distribution where $k$ is some number of trials, and let $\Phi(B(P,k), x)$ denote the cumulative distribution function of the binomial up to $x$. The probability that at least one purification is successful is then 

\begin{equation}
    P_{\geq 1} \equiv 1 - \Phi(B(P,k), 1)
\end{equation}

The minimum number of circuits needed to achieve an overall confidence $S$ is therefore

\begin{equation}
    K \equiv \min_k  (P_{\geq 1} \geq S)
\end{equation}

The purification factor is now the number of multiplexed circuits times the number of pairs needed for each circuit.

\begin{equation}
    \chi = K \times 2^{N}
\end{equation}


Another important consideration for satellite based entanglement distribution is how the entanglement ought to be encoded into the photon pairs.
One established and robust option is the polarization basis.
Corroborating theoretical and experimental results indicate that polarization mode errors are comparatively small for atmospheric transmission \cite{Bonato_2009}.
For our study we let $F_0 = 0.87$ which is the collection fidelity reported by Quantum Experiments at Space Scale (QUESS) group \cite{Mozi}.
For the sake of argument we set our target fidelity to be $F_{id} = 0.999$, and our confidence rating as $S = 0.999$.
With these parameters we find that $N = 2$ rounds of purification are sufficient to meet the target fidelity.
This has a corresponding success rate of $P = 0.573$ which indicates $K = 9$ and $\chi = 36$.






\section{Atmospheric and free-space attenuation}

Let $\eta$ be the double down-link attenuation of a photon pair from a satellite or in other words, the success rate of pair transmission.
Our objective now is to determine the rate of photon pairs required to meet the $R_{IP+P}$ rate needed for purification and lattice surgery, again with some confidence $S$.
The number of pairs that reach the ground after $k$ attempts is a random variable that follows a binomial distribution, however since the attempt rate is very large, $(k \gg 1)$ we approximate this as a Normal distribution with mean $k\eta$ and variance $k\eta(1-\eta)$

\begin{equation}
    \mathcal{N}(k\eta, \; k\eta(1-\eta))
\end{equation}

Similar to our calculations for the purification factor, the probability that at least $R_{IP+P}$ photon pairs are transmitted given $k$ attempts is the upper area of the probability density function from the point $R_{IP+P}$.

\begin{equation}
P_{\geq R_{IP + P}} \equiv 1 - \Phi(\mathcal{N}(k\eta, \; k\eta(1-\eta)), R_{IP + P})
\end{equation}

The required pair generation rate of the satellite, $R_{PG}$ is therefore the minimum value of $k$ such that the probability of exceeding $R_{IP+P}$ is greater than or equal to the success rate:

\begin{equation}
    R_{PG} = \min_k (P_{\geq R_{IP + P}})
\end{equation}

A further simplification is possible using Markov's inequality, which gives an upper bound for the probability that a random variable $X$ of some distribution is greater than a constant $a$

\begin{equation}
P(X \geq a) \leq \frac{E(X)}{a}
\end{equation}

Adapting this inequality for our case by setting $X \rightarrow R_{PG}$, $a \rightarrow R_{IP+P}$, and $E(X)  \rightarrow R_{PG} \times \eta$, we find that

\begin{equation}
S = P(R_{PG} \geq R_{IP+P}) \leq \frac{R_{PG} \; \eta}{R_{IP+P}}
\end{equation}

Therefore if $S = 1-\epsilon$ where $0 < \epsilon \ll 1$, we can approximate the pair generation rate as

\begin{equation} \label{eq:rpg}
R_{PG} \approx \frac{R_{IP+P}}{\eta}
\end{equation}

Substituting with equations \ref{eq:R_IP+P} and \ref{eq:R_IP} respectively, we momentarily conclude by obtaining an expression that relates the photon pair rate to the logical pair rate

\begin{equation}
R_{PG} = \frac{D^2 R_{LP} \; \chi}{\eta}
\end{equation}

\begin{table*}[ht!]
\begin{tabular}{ | c | c | c | } 
  \hline
  Classification  & City pairs & Average loss (dB)\\ 
  \hline
  State (500-999 km) & Toronto - New York City & 45.1 \\
  \hline
  Continental (1000-4999 km) & Sydney - Auckland & 65.6 \\
  \hline
  Transcontinental (5000 km+) & New York City - London & 79.1 \\
  \hline
\end{tabular}
\caption{\label{tab:cityloss} Most optimistic average double down-link losses for three distance classifications over a 24 hour period for a 400 satellite constellation \cite{Khatri_2021}.}
\end{table*}
Let us now focus our attention on choosing a suitable value for $\eta$.
Much work has been done to characterise atmospheric attenuation for satellite based entanglement distribution.
Bonato et.\ al.\ developed a model for $\eta$ in the context of quantum key distribution (QKD) \cite{Bonato_2009} as did Mazzarella et.\ al.\ \cite{Mazzarella20} and Khatri et.\ al.\ \cite{Khatri_2021}.
The results of Khatri et.\ al.\ are especially relevant for our discussion, as they demonstrated the feasibility of a quantum satellite network for QKD and conducted extensive simulations on a 400 satellite constellation to find average down-link attenuations between major cities.
We elected to use their three most optimistic attenuation rates for state, continental and transcontinental distances, the values of which are presented in table \ref{tab:cityloss}.
Due to the importance of these results in our calculations, a brief overview of their attenuation model and satellite constellation is provided for completeness.

The major contribution in down-link attenuation is attributed to beam widening.
This has contributions from natural free-space widening and atmospheric diffraction.
Although atmospheric diffraction contributes a higher widening rate, the team counter-intuitively demonstrates that free-space widening is the more significant of the two contributions.
This is because the atmospheric depth over which diffraction takes place is considerably smaller than the free-space distance.
Beam wandering, where the median of the Gaussian profile shifts in the $(x,y)$ plane is known to be negligible for down-link transmission and so is not considered.
Perfect weather conditions are assumed, and the atmosphere is taken to be a homogeneous layer of constant density.
The effects of ambient light are considered on the daytime regions of the globe since sunlight contributes a significantly higher signal to noise ratio and therefore decreases transmission efficiency.

The design objective of the constellation is to supply continuous global coverage with as few satellites as possible such that no double down-link channel ever exceeds 90dB.
The constellation consists of a number of equally spaced rings of satellites in polar orbits with an identical number of satellites in each ring.
The optimal satellite configuration was decided as follows:
In the worst case scenario, two ground stations each located on the equator are separated by some distance $d$.
A constellation of 400 satellites was proposed which optimises their figure of merit: The ratio of average entanglement distribution to the total number of satellites.

\section{Required Pair Generation Rate and Satellite Power}

In this section, we relate the pair generation rate $R_{PG}$ to the required satellite power $P_s$.
We assume all power is exclusively allocated to the task of pair production.
The rate at which a satellite can generate pairs $(S_{PG})$ 
depends on the brightness of the source $(N_p)$, the power consumption of each source $(P_r)$, and the power available to the satellite $(P_s)$.
    
\begin{equation}
    S_{PG} = \frac{P_s N_p}{P_r}
\end{equation}

Rearranging this, and setting $S_{PG} \rightarrow R_{PG} $ gives us the required satellite power for a particular generation rate:

\begin{equation}
    P_s = \frac{R_{PG} P_r}{N_p}
\end{equation}

Substituting $R_{PG}$ with eq.\ \ref{eq:rpg}, we obtain a closed form expression relating the satellite power to the rate of logical pair production.

\begin{equation} \label{eq:Ps}
    P_s = \frac{D^2 R_{LP} \; \chi P_r}{N_p \eta}
\end{equation}

To determine a realistic upper bound for the maximum satellite power, we present a brief survey of the power ratings of Indian communication satellites in table \ref{tab:sat}.
The most powerful of these, the GSAT-11, has a considerably larger power rating than the others but at a comparable budget.
We therefore estimate that a satellite with a power rating of $10$ kW is on the order of the most powerful commercial satellite possible with current technology.

\begin{table}
\begin{tabular}{ | c | c | c |  } 
  \hline
  Satellite Name & Power & Budget (10 million USD) \\ 
  \hline
  GSAT-11 & \href{https://www.isro.gov.in/gsat-11-mission/gsat-11-mission-brochure}{13.6 kW} & \href{http://164.100.24.220/loksabhaquestions/annex/176/AS232.pdf}{7.43} \\ 
  \hline
  GSAT-31 & \href{https://www.isro.gov.in/gsat-31/gsat-31-brochure}{4.7 kW} & \href{http://164.100.24.220/loksabhaquestions/annex/176/AS232.pdf}{6.46} \\ 
  \hline
  GSAT-7A &\href{https://www.isro.gov.in/gslv-f11-gsat-7a-mission/gslv-f11-gsat-7a-brochure}{3.3 kW} & \href{https://timesofindia.indiatimes.com/india/why-isros-gsat-7a-launch-is-important-for-iaf/articleshow/67153347.cms}{6.32-10.11} \\
  \hline
  GSAT-29 & \href{https://www.isro.gov.in/gslv-mk-iii-d2-gsat-29-mission/gslv-mk-iii-d2-gsat-29-mission-brochure}{4.6 kW} & \href{http://164.100.24.220/loksabhaquestions/annex/176/AS232.pdf}{2.08} \\
  \hline
  GSAT-30 & \href{https://www.isro.gov.in/gsat-30/gsat-30-curtain-raiser-video-english}{6 kW} & \href{http://164.100.24.220/loksabhaquestions/annex/176/AS232.pdf}{6.46} \\
  \hline
\end{tabular}
\caption{\label{tab:sat} A survey of Indian communication satellites launched between 2018 and 2019 with power ratings and costs}
\end{table}

The brightest available Bell-pair source reported at the time of writing is a waveguide integrated AlGaAson micro-resonator with a brightness of $20 \times 10^9$ Bell pairs $\textrm{s}^{-1} \textrm{mW}^{-2}$ \cite{Steiner21}.
Its high output combined with a micrometer scale form factor makes it a promising candidate as a satellite-based entanglement source.
From the experimental data, the highest attainable rate reported was $4 \times 10^6$ pairs per second at a power of $15 \mu W$.
According to the team, increasing the power beyond this point would exceed the lasing threshold of the micro-resonator which would reduce the overall entanglement visibility.
With this information, we set $N_P$ and $P_r$ to the aforementioned values of brightness and power per source respectively.


\section{Results and Discussion} \label{sec:results}

\begin{figure*}[ht]
    \includegraphics[width=\textwidth]{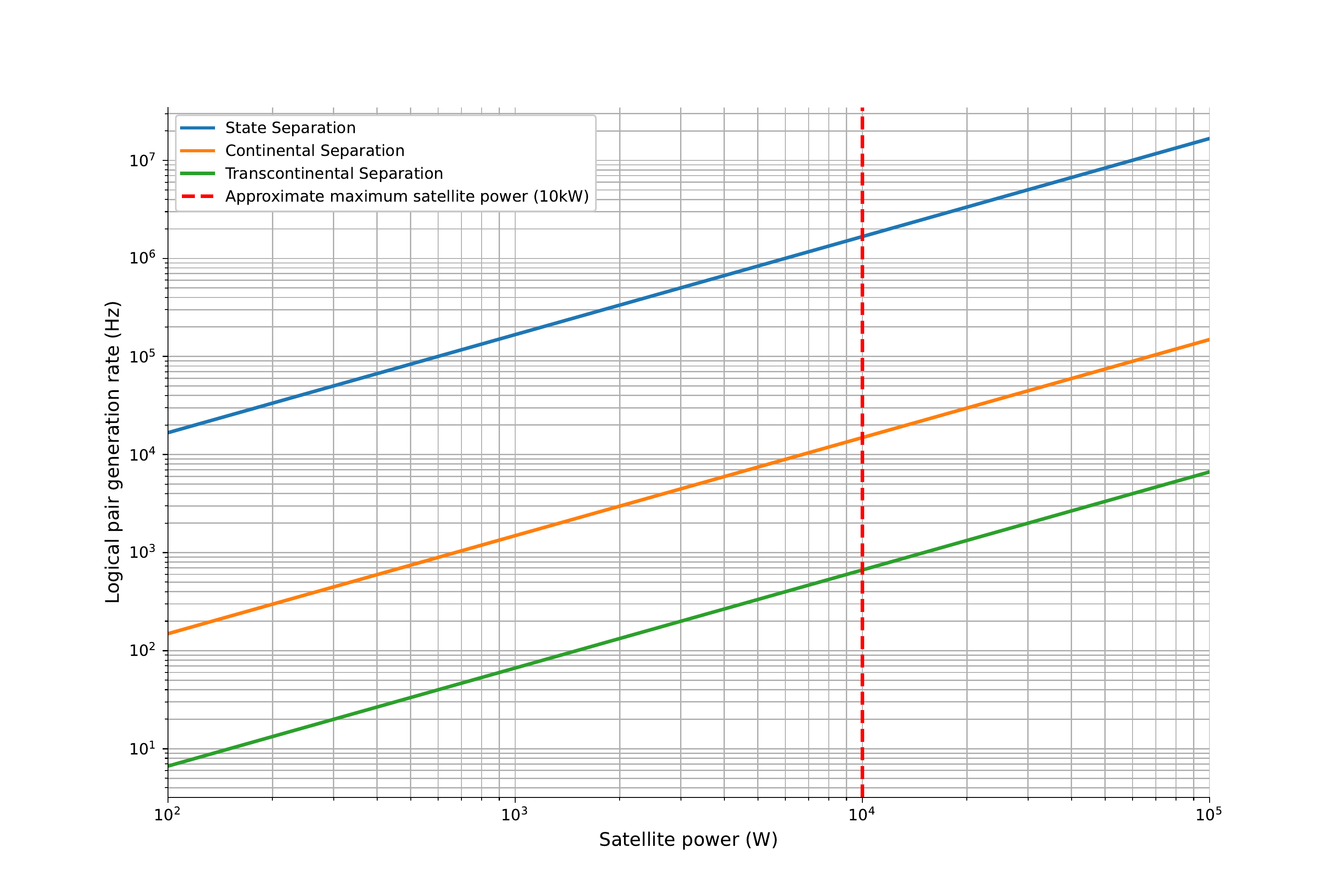}
    \caption{\label{fig:main} The rate at which logical surface code Bell pairs can be generated versus the available satellite power (eq.\ \ref{eq:Ps}) for three different distance ratings (Table \ref{tab:cityloss}).
    The vertical dashed line indicates the approximate maximum power of a commercial satellite.
    }
\end{figure*}

Let us begin by discussing the significance of estimating maximum achievable logical pairs rates in the context of distributed quantum computation.
The rate at which logical Bell pairs can be generated is essentially the \textit{global clock-speed} of a distributed quantum computer.
More precisely, the logical operations of state teleportation, gate teleportation, and non-local two qubit operations are all rate-limited by logical pair production.
By estimating the maximum possible rate of logical pairs, we indicate the rate at which \textit{non-local} operations between distant qubits can be performed.

How does the global clock-speed of a distributed quantum computer relate to its overall utility?
This is a difficult (if not impossible) question to answer in the absolute sense, but in general we understand that a fast quantum computer is preferable over a slow one.
Let us propose a thought experiment to resolve this insight with greater resolution.
Suppose we have a quantum computer $q$ and a quantum algorithm $j$ that takes an integer as input.
Let $J$ be a fictitious oracle that takes a clock-speed as input and returns the smallest integer such that the quantum algorithm $j$ achieves supremacy on $q$ (i.e.\ completes the computation faster than any existing classical computer).
We expect that $J$ is a continuous function since in principle supremacy is possible at any clock-rate (provided one chooses a sufficiently big input), and we expect that $J$ is monotonically increasing since it would be absurd for a slow quantum computer to achieve supremacy before a fast one.
From these properties of $J$, we see that \textit{fast quantum computers can reach supremacy with smaller computational problems than slow ones.}
This means that fast quantum computers are likely to be useful for a \textit{broad range} of problems, whereas slow quantum computers will only realise an advantage for very large calculations.
Additionally, a slow quantum computer will require more physical resources than a fast one in order to achieve supremacy.
It is for these two reasons that fast quantum computers are \textit{strongly preferred} over slow ones.
Although we cannot quantify the utility of a distributed quantum computer given its global clock-speed, we can at least compare our estimates to the average gate times of a variety of \textit{physical qubits} (Table \ref{tab:gatetimes}).
In this way, we get an idea of how powerful our hypothetical distributed system is given our current understanding of what's possible with these contemporary systems.

\begin{table}
\begin{tabular}{|c|c|c|} 
\hline
Architecture & Average Gate Time & Rate \\ [0.5ex]
\hline
Superconducting qubits \cite{noguchi2020fast} & 50 $ns$ & $2 \times 10^7$ Hz \\
\hline
 NV Diamond \cite{chou2015optimal} & 0.05 $\mu s$ & $2 \times 10^7$ Hz \\
\hline
Ion trap \cite{schafer2018fast} & 1.6 $\mu s$ & $6.25 \times 10^5$ Hz \\ 
\hline
NMR Spins \cite{nmr} & $~1 ms$ & $1 \times 10^3$ Hz \\
\hline
\end{tabular}

\vspace{0.5cm}

\begin{tabular}{|c|c|}
    \hline
     Distance Category & Rate \\
     \hline
     State & $2 \times 10^6$ \\
     \hline
     Continental & $1 \times 10^4$ \\
     \hline
     Transcontinental & $6 \times 10^2$ \\
     \hline
\end{tabular}

\caption{\label{tab:gatetimes} Sample of average gate times for common qubit architectures}
\end{table}

Our numerical results are presented in figure \ref{fig:main}.
Here, we plot the generation rate of logical Bell pairs (as given by eq.\ \ref{eq:Ps}) versus the required satellite power for state, continental, and transcontinental distance ratings (Table \ref{tab:cityloss}).
Additionally, we plot a vertical dashed line indicating the approximate maximum power of a commercial satellite ($10$kW).
We estimate the fastest possible logical pair rates for each distance category by looking at where the three curves intersect this vertical.
For state distances, this rate is around $2 \times 10^{6} \; \textrm{s}^{-1}$. For continental distances, $\approx 1 \times 10^{4} \; \textrm{s}^{-1}$, and for transcontinental distances, $\approx 6 \times 10^{2} \; \textrm{s}^{-1}$.
Let us compare these estimates with the average gate times of common qubit technologies (Table \ref{tab:gatetimes}).
Here we see that the maximum achievable rate at the statewide distance is comparable to the rate of trapped-ion systems.
Continental and transcontinental distances in turn are comparable to the speeds of an NMR-spin quantum computer.

With this in mind, we stress that these upper bounds are \textit{far from realistically achievable} due to the numerous highly optimistic parameter assumptions and simplifications we made throughout this work.
We treated photon capture and conversion as a lossless process, and assumed that quantum memories and local operations are effectively noiseless.
We treated incoming pairs with a special noise model to improve the purification rate.
Our adapted model for double-down link attenuation assumes ideal weather conditions, and we assume that $100\%$ of a satellites power can be allocated exclusively to photon pair production.
We also point out that all resource estimation up to this point has been done with respect to a single pair of distributed surface codes, which suggests the communication infrastructure would need to scale in proportion to the number of distributed qubits.

Let us now consider the ways in which we might improve the performance of a quantum satellite network.
Our options are to increase the throughput of the satellites, decrease the required pair generation rate, improve the efficiency of our purification or reduce the relative attenuation of the down-link channel.
In the first case, the only possibilities are to increase the satellite power or to improve the brightness of the photon pair source.
It is unlikely that the power available to a commercial satellite will dramatically increase, though we suspect the brightness of entanglement sources will be improved at least an order of magnitude in the near-distant future.
With respect to the pair generation rate (eq: \ref{eq:rpg}), we find that the code distance contributes one order of magnitude and is almost certainly irreducible for the surface code since error-corrected logical qubits are required for distributed quantum computation.
It is possible however that choosing another code-types may yield a small advantage.
In our work, we considered a parity check recurrence purification and showed the corresponding purification factor contributes one order of magnitude to the required pair rate.
Given our optimistic noise assumptions, we do not consider it likely that our estimate of $\chi = 36$ will be dramatically improved, though this remains an open question.

It was brought to our attention that it may be possible to avoid non-deterministic recurrence purification by using a pair generation protocol that is different from the one depicted in Fig.\ \ref{fig:spacetime}.
Here, the idea is to implement an error correcting code over multiple \textit{imperfect} logical Bell states, which would eliminate the need to purify initial entanglement resources.
For a specific example, we extend our gratitude to Craig Gidney for providing us with a Stim implementation of a surface-code purification using a 5-qubit code \cite{Gidney}.
One drawback with this approach however is that concatenating error-correcting codes increases the length of the space-time circuit which in turn decreases the rate at which logical pairs can be generated.
A promising direction for further research is whether or not such a strategy could yield an advantage for logical pair production.

By far the most significant contribution to the required rate of photon pairs is the double down-link attenuation.
The optimistic results we selected from Khatri et.\ al.\ contribute between five and eight orders of magnitude depending on the distance between stations.
As this loss results from free-space transmission and atmospheric diffraction, there are few options for mitigating this effect.
One notable strategy is to store half of a pair on the satellite as the other transmits rather than sending both halves at the same time.
We credit Alexis Shaw with this idea.
At first this seems like only a superficial difference, but it turns out that this trick can effectively \textit{halve the attenuation rate}.
This is because when half of a pair is established on the ground, the other half can be used for \textit{entanglement swapping} with another established pair (albeit with a success probability of 50\%).
A significant disadvantage with this approach though is that the satellite must reliably control, process and measure an enormous number of pairs which is presently infeasible for a satellite.



Up to this point we considered a down-link transmission model where entanglement is generated in satellites and distributed between ground stations.
The reverse case is up-link transmission where entanglement pairs are prepared on the ground and fired up to a satellite which performs entanglement swapping to project the pair between the stations.
The main advantage of this approach is that ground stations can generate significantly more photon pairs since power is no longer a major limiting factor.
The downside however is that the attenuation rate for up-link is significantly higher than for down-link. 
This is because beam wandering effects from atmospheric turbulence are more significant when the Gaussian profile is small.
Engineering challenges are another difficulty for these hypothetical networks.
Unlike down-link satellites which are relatively passive, the up-link satellite would be required to control and measure incoming photon pairs with quantum precision.
This is a demanding task even for a laboratory on Earth.
Whether or not these trade-offs are overall beneficial remains to be seen, though our group is currently investigating this is greater detail.

We believe our results indicate a need to reconsider the problem of long-distance entanglement distribution.
One understated consideration of quantum networks is that they have \textit{no latency requirements}.
Unlike classical data networks, entanglement can be stored as a physical resource and transported by moving error corrected memory units.
This is the motivating principle of the \textit{quantum sneakernet} \cite{sneakernet}, which may be a more viable long-term alternative to quantum satellite networks.
We note that the quantum sneakernet has its own significant engineering challenges since it's predicated on the aspiration it's possible to make a sufficiently portable and scalable quantum memory unit that can be transported long distances.
Given the alternatives however, we feel that the sneakernet has a greater potential for scalability than any quantum satellite network.
This is because satellites will always be rate limited by power consumption which is not a problem in the sneakernet model.


In this work, we determined upper bounds for
the rates of logical pair generation between surface code qubits at a variety of distances when the physical entanglement is supplied by a quantum satellite network.
We began by establishing a reasonable code distance that was predicated on an arbitrarily ``hard'' computational problem.
We accounted for losses in entanglement purification by considering a $2 \rightarrow 1$ parity check recurrence protocol under an optimal noise model.
In order to calculate our estimates for attainable pair rates, we also developed a closed form equation relating the available satellite power to the achievable logical pair production rate.


\section{Acknowledgements}

The authors wish to thank Katanya Kuntz, Thomas Jennewein, and Pablo Avila for their feedback and insights -- in particular for introducing us to the possibility of up-link satellite networks as an avenue for further research.
We thank Craig Gidney for his suggestion on avoiding non-deterministic recurrence purification and providing us with a proof of concept \cite{Gidney}.
Finally, the student authors wish to thank our colleagues and supervisors for attending our seminar regarding this paper at the UTS Centre for Quantum Software and Engineering.

This research was conducted by the Australian Research Council Centre of Excellence for Engineered Quantum Systems (project CE170100009) and funded by the Australian Government.
The views, opinions and/or findings expressed are those of the author(s) and should not be interpreted as representing the official views or policies of the Department of Defense or the U.S. Government.  This research was developed with funding from the Defense Advanced Research Projects Agency [under the Quantum Benchmarking (QB) program under award no. HR00112230007 and HR001121S0026 contracts].
H.\ L.\ was supported by the ARC Centre of Excellence for Quantum Computation and Communication Technology (CQC2T), project number CE170100012

\vspace{0.25cm}
\begin{center}
    A.M.D.G.
\end{center}

\section{Appendix -- tables of constants}

\begin{table}[h]
\begin{tabular}{|c | c | c|}
\hline
 Parameter & Value & Justification \\ [0.5ex] 
\hline
$D$ & 37 & Required code distance for FT RSA \\
\hline
$F_0$ & 0.87 & QUESS Satellite data \cite{Mozi} \\
\hline
$F_{\textrm{id}}$ & 0.999 & Required fidelity rating \\
\hline
$S$ & 0.999 & Requirement of surface code \\
\hline 
$\chi$ & 36 & Best estimate \\
\hline
$P_s$ & 10 kW & Highest power rating for comms sat. \\
\hline
$P_r$ & 15 $\mu$W & Power of Brightest Bell source \cite{Steiner21}\\
\hline
$N_P$ & 4 $\times 10^6$ $s^{-1}$ & Throughput of Bell source \cite{Steiner21}\\
\hline
\end{tabular}
\caption{\label{table:params} Summary of selected parameter values with justifications}
\end{table}

\begin{table}[h!]
    \centering
    \begin{tabular}{|c|p{6cm}|} 
    \hline
    Constant & Definition \\
    \hline
    $D$ & Surface code distance \\
    \hline
    $R_{LP}$ & Logical pair generation rate \\
    \hline
    $R_{IP}$ & Rate of ideal bell pairs required to produce logical pairs at rate $R_{LP}$ \\
    \hline
    $P_{LB}$ & Tolerable error rate of logical pair production \\
    \hline
    $S$ & The likelihood that a given routine succeeds (also called the confidence). \\
    \hline
    $\chi$ & Number of imperfect pairs needed to purify down to one ideal pair with confidence $S$ \\
    \hline
    $R_{PG}$ & Required photon pair rate to generate logical pairs at rate $R_{LP}$ \\
    \hline
    $\eta$ & Double down-link attenuation rate (The success rate of a photon-pair transmitting) \\
    \hline
    $P_s$ & Satellite power \\
    \hline
    $P_r$ & Power consumption per photon pair source \\
    \hline
    $N_P$ & Brightness of Bell source (pairs per unit time) \\
    \hline
    \end{tabular}
    \caption{Table of the most significant constants with definitions}
    \label{tab:constants}
\end{table}

\bibliography{paperbib}

\end{document}